\documentclass[12pt]{article}
\usepackage{epsfig}
\usepackage{psfrag}
\usepackage{latexsym}
\def\beq{\begin{equation}}
\def\eeq{\end{equation}}
\def\bea{\begin{eqnarray}}
\def\eea{\end{eqnarray}}
\def\bet{\begin{tabular}}
\def\eet{\end{tabular}}
\def\bes{\begin{subequations}\bea}
\def\ees{\eea\end{subequations}}

\newcommand{\bi}{\begin{itemize}}
\newcommand{\ei}{\end{itemize}}
\def\be{\begin{equation}}
\def\ee{\end{equation}}
\def\bc{\begin{center}}
\def\ec{\end{center}}
\def\bea{\begin{eqnarray}}
\def\eea{\end{eqnarray}}

\begin{document}
\begin{titlepage}
\vspace*{-1cm}
\phantom{hep-ph/******} 
\hfill{ }
 
\begin{flushright}
{CERN-PH-TH/2010-225 \\
EURONU-WP6-10-26}
\end{flushright}
\vskip 2.5cm
\begin{center}
\mathversion{bold}
{\Large\bf Importance of nuclear effects in the measurement of neutrino oscillation parameters}
\mathversion{normal}
\end{center}
\vskip 0.2  cm
\vskip 0.5  cm
\begin{center}
{\large Enrique Fernandez Martinez}~\footnote{e-mail address:enrique.fernandez.martinez@cern.ch}\\
Max-Planck-Institut f\"ur Physik
(Werner-Heisenberg-Institut),\\ F\"ohringer Ring 6, 80805 M\"unchen,
Germany and CERN, PH-TH Division, CH-1211 Gen\`eve 23, Switzerland.\\
\vskip .1cm
\vskip .2cm
{\large Davide Meloni}~\footnote{e-mail address:davide.meloni@physik.uni-wuerzburg.de}
\\
\vskip .1cm
 Institut f{\"u}r Theoretische Physik und Astrophysik, \\
Universit{\"a}t W{\"u}rzburg, D-97074 W{\"u}rzburg, Germany
\\ 
 
\end{center}
\vskip 0.7cm
\begin{abstract}
We investigate how models for neutrino-nucleus cross sections based on different assumptions for the nuclear dynamics 
affect the forecasted sensitivities to neutrino oscillation parameters at future neutrino facilities. 
We limit ourselves to the quasi-elastic regime, where the neutrino cross sections can be evaluated with less
uncertainties, and discuss the sensitivity reach to $\theta_{13}$ and $\delta$ at a 
prototype low-$\gamma$ $\beta$-beam, mostly sensitive to the quasi-elastic regime.
\end{abstract}
\end{titlepage}
\setcounter{footnote}{0}
\vskip2truecm

\section{Introduction}

In recent years the experimental study of neutrino oscillations has much contributed to our knowledge of particle physics by establishing non vanishing neutrino masses and by measuring or constraining the 
corresponding mixing angles. Within the domain of neutrino oscillations, the main goal of the next generation of facilities 
is the measurement of the mixing angle $\theta_{13}$, which at present is only limited by an upper bound, and the observation of leptonic CP violation, for which we have no hints at the moment. 
These measurements are extremely challenging, since the smallness of the present bound on $\sin \theta_{13} < 0.22$ at $3 \sigma$ \cite{GonzalezGarcia:2007ib} constrains the signals to be very subleading. 
It is therefore necessary to have a good knowledge of the detector response function as well as an accurate knowledge of the beam to control the various systematic errors with a very good precision. 
For this task, novel $\nu$ beams with extremely low backgrounds and systematic uncertainties have been proposed for future measurements based on the decay of muons (Neutrino factories~\cite{Geer:1997iz,De Rujula:1998hd}) or $\beta$ unstable ions 
($\beta$-Beams~\cite{Zucchelli:sa}). The $\nu_e$ fluxes from such decays would be extremely pure and the systematic errors very small, especially if the flux is normalized through a measurement with a near detector. 
Conversely, the systematic uncertainties associated with the detection process should be kept at the same accuracy level. It is usually argued that the $\nu_\mu/\nu_e$ neutrino-nucleus cross section ratio can be 
understood at the level of some percent and that the $\nu_e$ cross section can be measured with similar precision at a near detector so that the $\nu_\mu$ signal can have systematic uncertainties of the order of 
a few percent at the far detector. This, however, does not take into account our present ignorance of the actual value of the cross section at low energies and how different models lead to forecasted sensitivities 
to the unknown parameters that differ by much larger margins than the few percent uncertainty normally considered. This question can severely impact the comparison of the relative performance of different 
facilities depending on the model adopted to parametrize their cross sections, particularly if the different facilities are sensitive to distinct energy regions. In this letter we want to discuss more in 
detail this question showing that, when a realistic model of nuclear dynamics is adopted, the neutrino cross section can sizably affect the forecasted precision measurement of $\theta_{13}$ and the
CP violating phase $\delta$. It is beyond the scope of this letter to discuss (and critically revise) all possible models for neutrino interactions, 
so we restrict ourselves to the Relativistic Fermi Gas Model (RFG), variants of which are widely used in 
many MonteCarlo codes and, as examples of more refined calculations,
to the one based on the Spectral Function approach (SF), on the Relativistic Mean Field (RMF) approximation and on the Random Phase Approximation (RPA), also including the effects of multinucleon
contributions.
The main features of these  models are  briefly summarized in Sect.\ref{summary} whereas in 
Sect.\ref{results} we show how the measurement 
of $\theta_{13}$ and $\delta$ can be affected by the different descriptions of the nuclear dynamics. We draw our 
conclusions in Sect.\ref{concl}.

\section{Summary of the charged current neutrino-nucleus cross sections}
\label{summary}
We work in the quasi-elastic regime (QE), which is of interest in many current and planned experiments (among them, MiniBooNE has already released its first cross section measurement \cite{AguilarArevalo:2010zc}).
 The doubly-differential cross section, in which a neutrino carrying initial four-momentum $k=(E_\nu,\bf k)$ scatters off a nuclear target to a state of four-momentum
$k^{'}=(E_\ell,\bf k^{'})$ can be written in Born approximation as follows:
\bea
\label{elem}
\frac{d^2\sigma}{d\Omega dE_\ell}=\frac{G_F^2\,V^2_{ud}}{16\,\pi^2}\,
\frac{|\bf k^{'}|}{|\bf k|}\,L_{\mu\nu}\, W_A^{\mu\nu} \ , 
\eea
where $G_F$ is the Fermi constant and $V_{ud}$ is the CKM matrix element coupling $u$
and $d$ quarks. The leptonic tensor, that can be written in the form 
\bea
\label{leptensor}
L_{\mu\nu}&=&8\,\left[k_\mu^{'}\,k_\nu+k_\nu^{'}\,k_\mu- g_{\mu\nu}(k\cdot
k^{'})-i\,\varepsilon_{\mu\nu\alpha\beta}\,k^{'\beta}\,k^\alpha \right]
\eea
is completely determined by lepton kinematics, whereas the nuclear tensor 
$W_A^{\mu\nu}$, containing all the information on strong interactions dynamics,
describes the response of the target nucleus. Its definition
involves the initial and final hadronic states $|0\rangle$ and $|X\rangle$,
carrying four momenta $p_0$ and $p_X$, respectively, as well as the nuclear
electroweak current operator $J^\mu_A$:
\bea
\label{hadronictensor}
W_A^{\mu\nu}&=& \sum_X \,\langle 0 | {J_A^\mu}^\dagger | X \rangle \,
      \langle X | J_A^\nu | 0 \rangle \;\delta^{(4)}(p_0 + q - p_X) \ , 
\eea
where the sum includes all hadronic final states.
The calculation of $W_A^{\mu\nu}$ is a complicated task which deserves some approximation; quite often the impulse approximation (IA) scheme is adopted, based on the
assumptions that at large enough ${\bf q}$ the target nucleus is seen by the
probe as a collection of individual nucleons and that the particles produced at
the interaction vertex and the recoiling $(A-1)$-nucleon system evolve
independently.
Within this picture, the nuclear current can be written as a sum of one-body
currents, i.e. $J^\mu_A \rightarrow  \sum_i \, J^\mu_i$, while the final state reduces to the direct product
of the hadronic state produced at the weak vertex (with momentum ${\bf p^{'}}$)
and that describing the $(A-1)$-nucleon residual system, with momentum $\bf p_{\cal R}$:
$| X \rangle \to | i,{\bf p}^{'} \rangle \otimes | {\cal R}, \bf p_{\cal R} \rangle$. 

\subsection{The Spectral Function approach}

The calculation of the weak tensor as described in Ref. \cite{Benhar:2005dj} naturally leads to the 
concept of Spectral Function. In fact, the final expression of the hadronic tensor can be cast in the following form:
\bea
W_A^{\mu\nu}&=& \frac{1}{2}\int d^3p\,dE \,P({\bf
p},E)\frac{1}{4\,E_{|\bf p|}\,E_{|\bf p+q|}} \,  W^{\mu\nu}(\tilde p,\tilde
q) \ ,
\label{hadtensor}
\eea
where $E_{\bf p}=\sqrt{|{\bf p}|^2+m_N^2}$ and the function $P({\bf p},E)$ is the target {\it Spectral Function}, i.e.
 the probability distribution of finding a nucleon with momentum 
${\bf p}$ and removal energy $E$ in the target nucleus. It then encodes all the 
informations about the
initial (struck) particle. The quantity $W^{\mu\nu}$ is the tensor describing
the weak interactions of the $i$-th nucleon in free space; the effect of
nuclear binding of the struck nucleon is accounted for by the
replacement $q=(\nu,{\bf q}) \to \tilde q=(\tilde \nu,{\bf q})$ 
with $\tilde \nu = E_{|\bf p+q|}-E_{|\bf p|}$. 
It follows that the second argument in the hadronic tensor is $\tilde p=(E_{|\bf p|},\bf p)$.
Substituting Eq.~(\ref{hadtensor}) into Eq.~(\ref{elem}), we get the final formula
for the {\it nuclear} cross section:
\bea
\label{final_cs}
\frac{d^2\sigma_{IA}}{d\Omega dE_\ell}=
\int d^3p\,dE \,P({\bf
p},E)\,\frac{d^2\sigma_{\rm elem}}{d\Omega dE_\ell} \ ,
\eea
in which we have redefined the {\it elementary} cross section as 
\bea
\label{elem_cs}
\frac{d^2\sigma_{\rm elem}}{d\Omega dE_\ell}=\frac{G_F^2\,V^2_{ud}}{32\,\pi^2}\,
\frac{|\bf k^{'}|}{|\bf k|}\,\frac{1}{4\,E_{\bf p}\,E_{|\bf p+q|}}\,L_{\mu\nu}
W^{\mu\nu} \ .
\eea
The calculation of $P({\bf p},E)$  has been only 
carried out for $A\leq4$ \cite{dieperink};
however, thanks to the simplifications associated with translation invariance, 
highly accurate results are also available for uniform nuclear matter, 
i.e. in the limit A $\rightarrow \infty$ with Z=A/2 \cite{bff} ($Z$ denotes
the number of protons).
The Spectral Functions for medium-heavy nuclei have been modeled using the
Local Density Approximation (LDA) \cite{bffs}, in which the experimental
information obtained from nucleon knock-out measurements is combined
with the results of theoretical calculations of the nuclear matter
$P({\bf p},E)$ at different densities.

\subsection{The Relativistic Fermi Gas}
The RFG \cite{Smith:1972xh} model, widely used in MonteCarlo simulations, provides the 
simplest form of the Spectral Function: 
\bea
\label{fermigas}
P_{RFGM}({\bf p},E)=\left(\frac{6\,\pi^2\,A}{p_F^3}\right)\,\theta(p_F-{\bf p})\,
\delta(E_{\bf p}-E_B+E) \ ,
\eea
where $p_F$ is the Fermi momentum and $E_B$ is the average binding energy,
introduced to account for nuclear binding. The term in parenthesis is a constant
needed to normalize the Spectral Function to the number of target nucleons, $A$.
Thus, in this model $p_F$ and $E_B$ are two parameters that are {\it adjusted} to 
reproduce the experimental data. For oxygen, the analysis of electron scattering 
data yields $p_F=225$ MeV and $E_B=25$ MeV \cite{Moniz:1971mt}.

\subsection{The Relativistic Mean Field approach}
Within the Relativistic Mean Field approximation we refer to the model described in \cite{Martinez:2005xe}, 
where, like in the previous cases, the nuclear current is written as a sum of single-nucleon
currents. The wave functions for the target and the residual nuclei are described in
terms of an independent-particle model. Then, the transition matrix elements can be cast
in the following form:
\begin{equation}
J^{\mu}_{N}(\omega,\vec{q})=\int\/\/ d\vec{p}\/
\bar{\psi}_F
(\vec{p}+\vec{q}) \hat{J}^\mu_N(\omega,\vec{q}\/) \psi_B(\vec{p})\; ,
\label{nucc}
\end{equation}
where  $\psi_B$ and $\psi_F$ are the wave functions for initial bound
and  final outgoing  nucleons, respectively, and $\hat{J}^\mu_N$ is
the relativistic current operator. In particular, the relativistic bound-state wave functions 
(for both initial and outgoing nucleons) are obtained as a solution of the Dirac equation,
in the presence of the same relativistic nuclear mean field potential, derived from a Lagrangian
containing $\sigma$, $\omega$ and $\rho$ mesons.
The calculated cross sections correctly account for the inclusive cross section which are interested in.

\subsection{The Random Phase Approximation}
The last model we want to take into account has been introduced in \cite{Martini:2009uj}, where the 
hadronic tensor is expressed in terms of the nuclear response functions 
treated in the  Random Phase Approximation (RPA). The response functions are related to the imaginary part of the 
corresponding full polarization propagators and the introduction of the RPA approximation means that 
the polarization propagators are the solutions of integral equations involving the bare propagators and 
the effective interaction between particle-hole excitations. Within this formalism, the authors of \cite{Martini:2009uj}
were able to show that multinucleon terms sizably increase the genuine charged current QE cross section  \cite{Martini:2009uj,Martini:2010ex}
in such a way to reproduce the MiniBooNE results \cite{AguilarArevalo:2010zc}.
The mechanism responsible for the enhancement that brings the theoretical cross section into agreement with the data
is multi nucleon knock out, leading to two particle-two hole (2p2h) nuclear final states. In the following, we will refer to
this ``generalized''  QE cross section as RPA-2p2h whereas we adopt the short RPA for the genuine QE cross section.

\subsection{Comparison of the cross sections}
\begin{figure}[h!]
\centering
\includegraphics[width=0.49\linewidth]{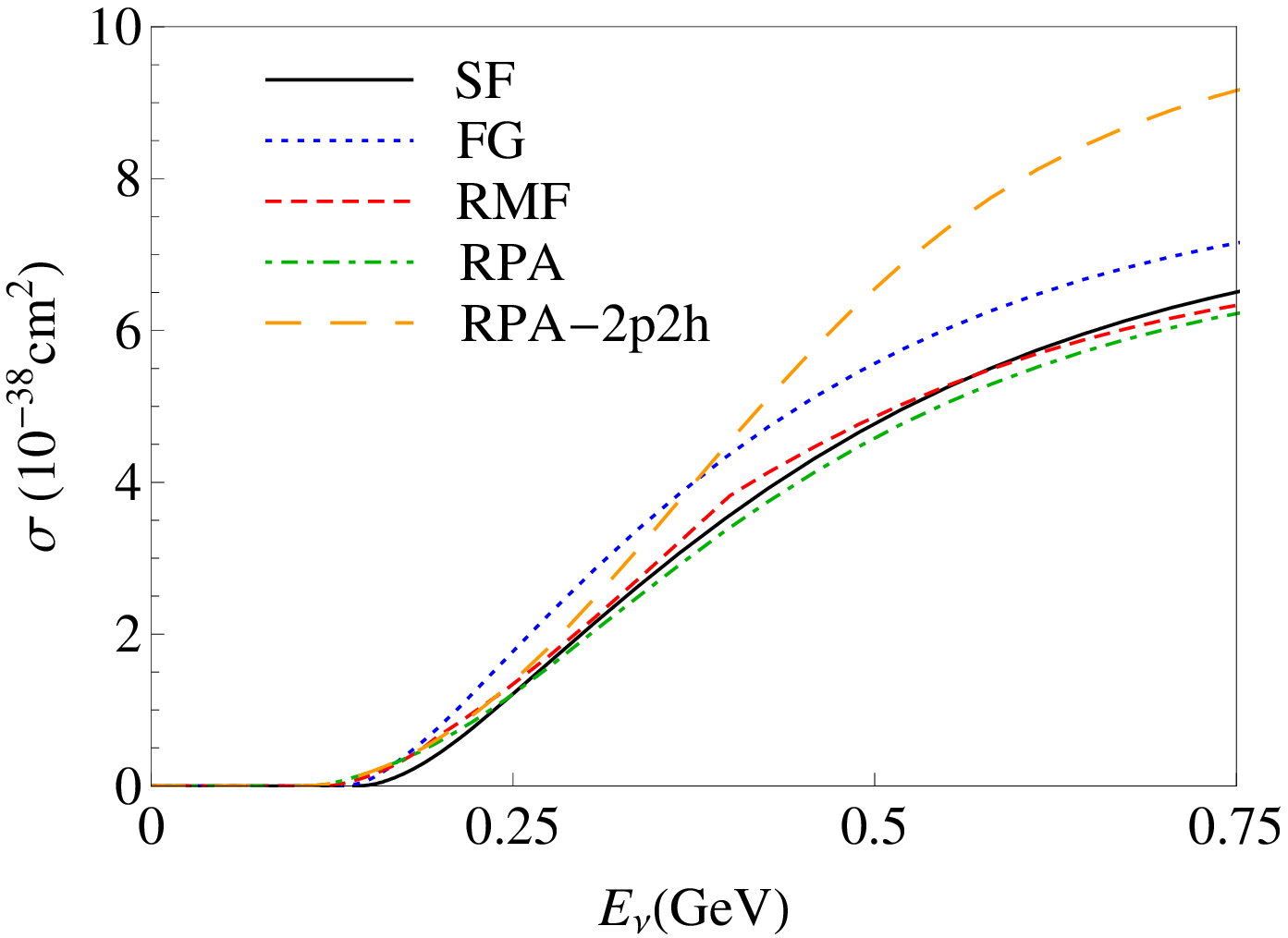}
\includegraphics[width=0.49\linewidth]{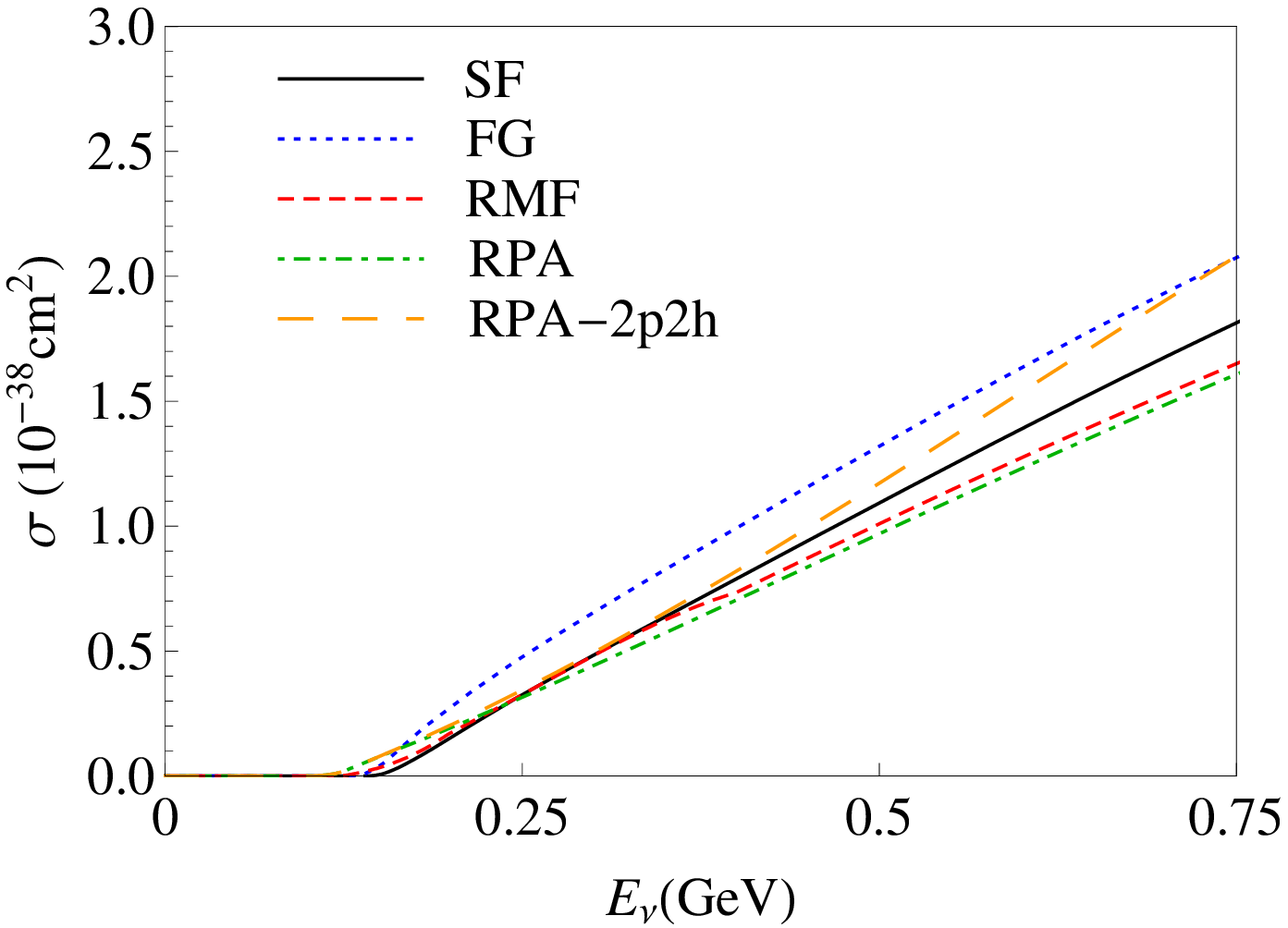}
\caption{\label{fig:xsect} \it Total charged current QE cross sections 
for the $\nu_\mu ^{16}O \to \mu^- X$ (left panel) and  $\bar \nu_\mu ^{16}O \to \mu^+ X$ (right panel) processes in the energy range $E_\nu\sim [0,0.75]$ GeV.}
\end{figure}

To summarize this section, we present in Fig.\ref{fig:xsect} a comparison of the five total QE cross sections 
for the $\nu_\mu \,^{16}O \to \mu^- X$ process (left panel) and $\bar \nu_\mu \,^{16}O \to \mu^+ X$ (right panel), in the energy range $E_\nu\sim [0,0.75]$ GeV. 
The curves have been computed using the dipole structure of the form factors and, in particular, a value of the axial mass
close to $m_A\sim 1$ GeV.
As it can be easily seen in the left panel, the RFG prediction sizably overestimates the SF, RMF and RPA results by roughly $15\%$, a fact that is well known to happen also for many other models with a 
more accurate description of the nuclear dynamics than the RFG approach (see, for instance, \cite{nuint09}). On the other hand, the inclusion of 2p2h contributions largely
enhances the QE cross section in the RPA approximation for energies above $\sim 0.5$ GeV, although it is smaller than the RFG at smaller energies.
For antineutrinos (right panel) the observed pattern is almost the same. 
From this comparison, a  {\it qualitative} understanding of the impact of different models of neutrino-nucleus cross sections on the forecasted sensitivity of future facilities can be already derived. 
A {\it quantitative} analysis of these effects will be the subject of the next section.

\section{The impact on the ($\theta_{13}$-$\delta$) measurement}
\label{results}
To estimate  the impact of different models of the cross section
on the measurement of $\theta_{13}$ and leptonic CP violation we choose a $\gamma = 100$
$\beta$-Beam facility as a representative example. The choice is motivated because the neutrino flux from such a facility spans up to $\sim 0.7$ GeV with the peak around $0.3$ GeV and is 
thus mostly sensitive to the quasielastic region explored here. The $\beta$-Beam concept was first introduced in Ref.~\cite{Zucchelli:sa} and involves  the production of $\beta$-unstable ions, accelerating 
them to some reference energy, and allowing them to decay in the straight section of a storage ring, resulting in a very intense and pure $\nu_e$ or $\bar \nu_e$ beams. The $\nu_e \to \nu_\mu$ ``golden channel'', 
which has been identified as the most sensitive to all the unknown parameters \cite{Cervera:2000kp},
can be probed at a far detector. We have considered here the original $\beta$-Beam proposal, 
where $^{18}$Ne ($^{6}$He) ions are accelerated to $\gamma \sim 100$ at the CERN SPS and stored so that $\nu_e$ ($\bar{\nu}_e$) beams are produced and the golden channel oscillation is searched for at a Mton 
class water Cerenkov detector located at $L=130$ km at the Frejus site, detailed analyses of the physics performance of this setup can be found in 
Refs. \cite{Mezzetto:2003ub,Mezzetto:2004gs,Donini:2004hu,Donini:2004iv,Donini:2005rn,Huber:2005jk,Campagne:2006yx,Winter:2008cn,Winter:2008dj,FernandezMartinez:2009hb}. 
In order to simulate the detector response when exposed to such a beam both in terms of signal efficiency and background, we have used the migration matrices derived in Ref.~\cite{Burguet-Castell:2005pa}. 
Systematic errors of $2.5 \%$ and $5 \%$ in the signal and background respectively have been taken into account. In all the simulations the following best fit values and $1 \sigma$ errors for the known oscillation 
parameters were assumed \cite{GonzalezGarcia:2007ib} 
$\Delta m^2_{21} = (7.6 \pm 0.2)\cdot 10^{-5}$ eV$^2$, $\Delta m^2_{31} = (2.5 \pm 0.1)\cdot 10^{-3}$ eV$^2$, $\theta_{12} = 34.0 \pm 1.0$ and $\theta_{23} = 45.0 \pm 3.6$. These parameters were marginalized 
over to present the final curves. The evaluation of the performance 
of the facility made use of the GLoBES software \cite{Huber:2004ka,Huber:2007ji}. It is important to notice that we are only using the quasielastic contribution to the neutrino cross section depicted in Fig.\ref{fig:xsect}. 

\begin{figure}[h!]
\centering
\includegraphics[width=0.458\linewidth]{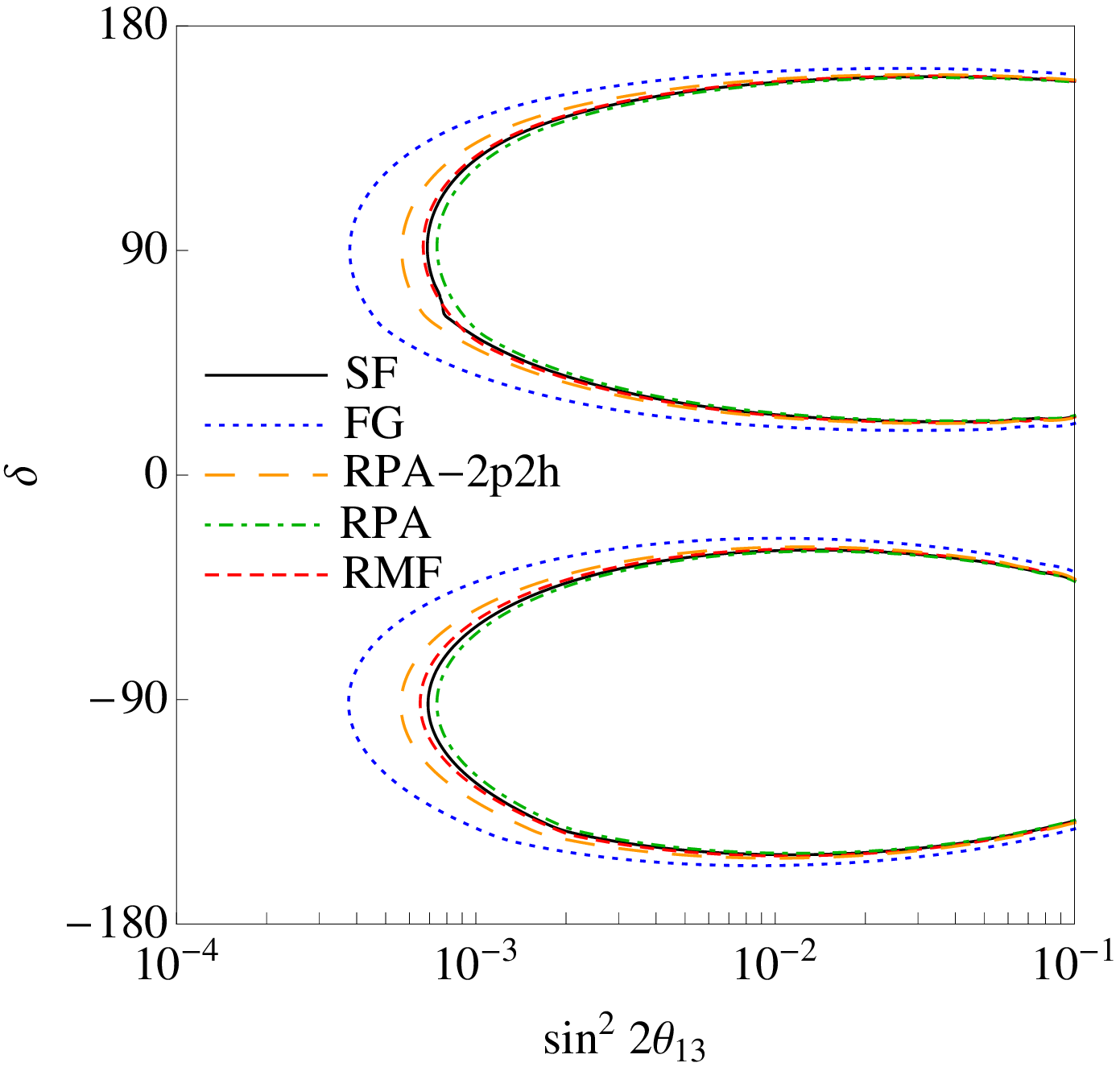}
\includegraphics[width=0.442\linewidth]{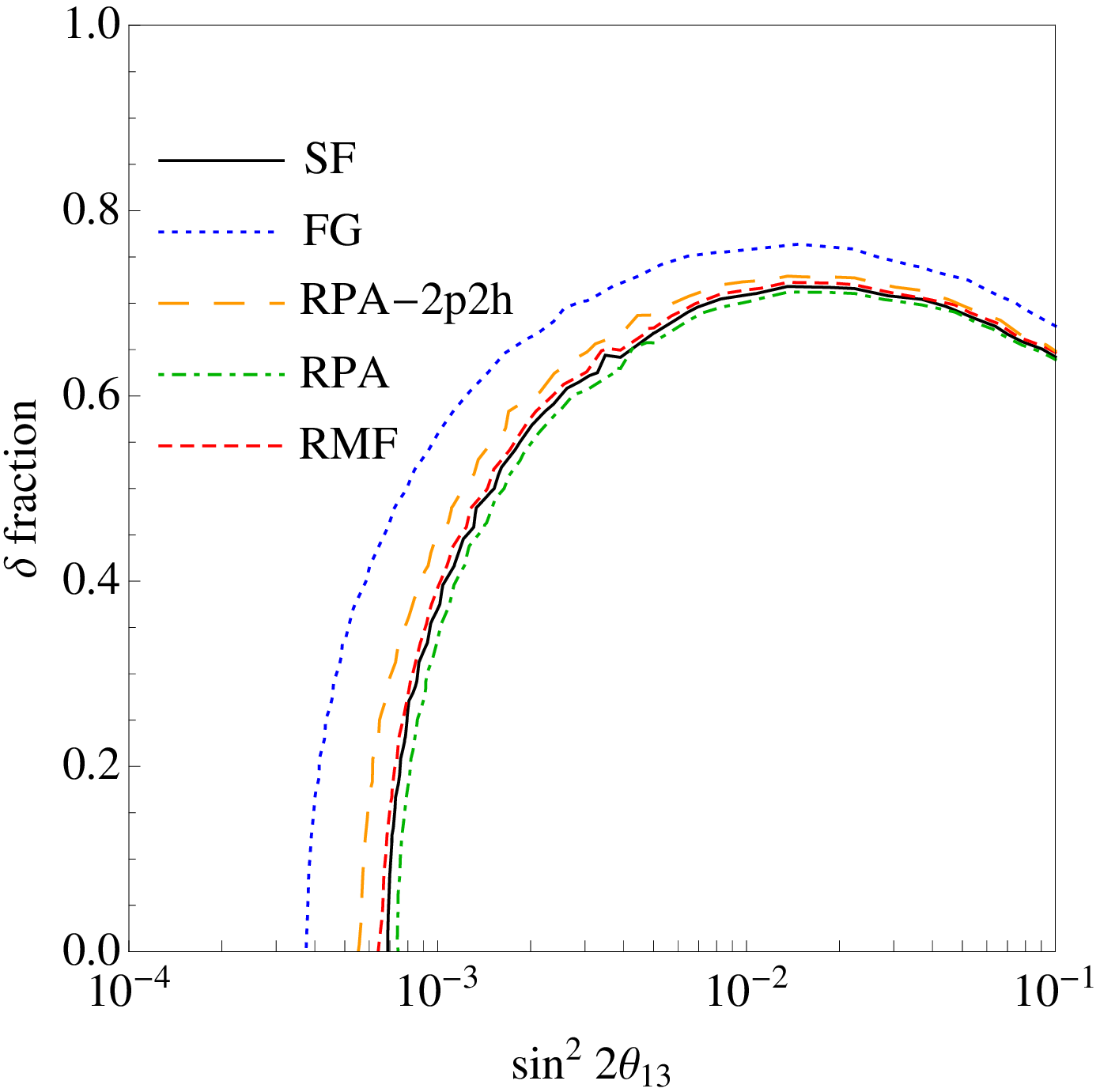}
\caption{\label{fig:disc} \it  Left panel: CP discovery potential in the $(\theta_{13},\delta_{CP})$ plane. 
Solid lines refer to the SF model, dotted lines to the RFG, short-dashed lines to the RMF, dot-dashed to RPA and long-dashed to RPA-2p2h.
Right panel: the CP fraction.}
\end{figure}

As an illustration we have focused on the dependence on the nuclear model adopted of two different observables, namely the {\it CP and $\theta_{13}$ discovery  potentials}, defined as the values of 
the CP-violating phase $\delta_{CP}$ and $\theta_{13}$ for which respectively the hypothesis of CP conservation $\delta_{CP}=0, \pm \pi$ or $\theta_{13}=0$ can be excluded at 3$\sigma$ after marginalizing over all other 
parameters. The CP discovery  potential is shown in Fig.\ref{fig:disc}, where we superimposed the results obtained using 
the RFG cross section and the SF, RMF and RPA calculations.
In the left panel, the CP discovery  potential is represented in the $(\theta_{13},\delta_{CP})$-plane; we clearly see that, for $\delta_{CP}\sim \pm 90^o$ (where the sensitivity is maximal) 
the RFG model gives a prediction which is around a factor 2 better than the SF, RMF and RPA models (which, as expected, behave almost in the same way) in $\sin^2 2\theta_{13}$ and around a 
$40\%$ better than the RPA-2p2h. This is not surprising because the $\beta$-Beam facility used in our simulations mainly probes energies smaller than $0.5$ GeV, where the RFG is still larger than any other model 
(see Fig.\ref{fig:xsect}). For the other points in the parameter space, the difference is less evident but still significant. 
The same information can be summarized in the right panel making use of the {\it $\delta$ fraction} ($\delta_F$), that represents the fraction of values of $\delta_{CP}$ for which CP can be discovered at a 
given $\theta_{13}$. The $\delta$ fraction has a maximum around $\sin^2 2\theta_{13}\sim 10^{-2}$ where the RFG model predicts $\delta_F \sim 0.75$ and the other models give $\delta_F \sim 0.7$.
\begin{figure}[h!]
\centering
\includegraphics[width=0.458\linewidth]{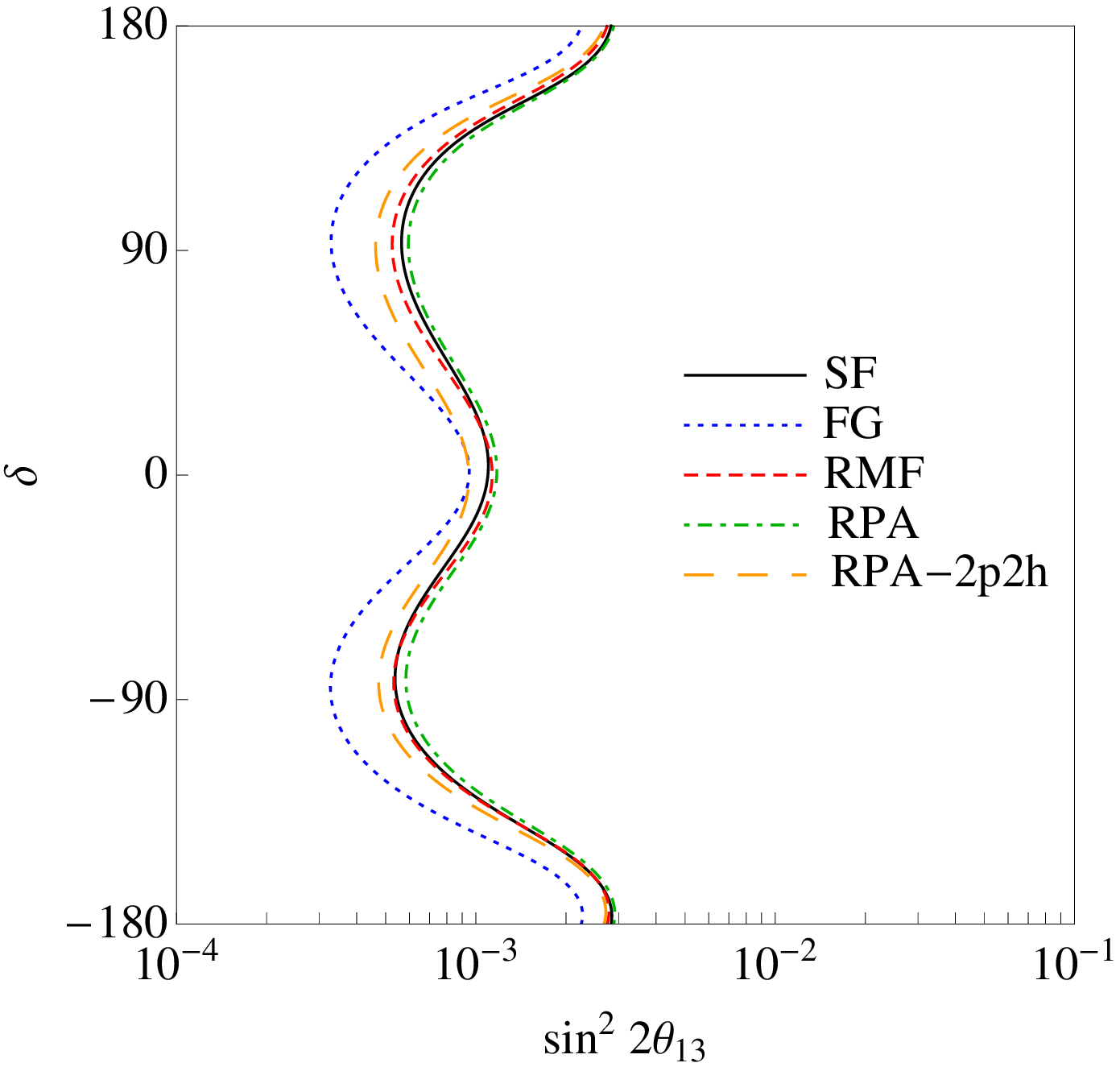}
\includegraphics[width=0.442\linewidth]{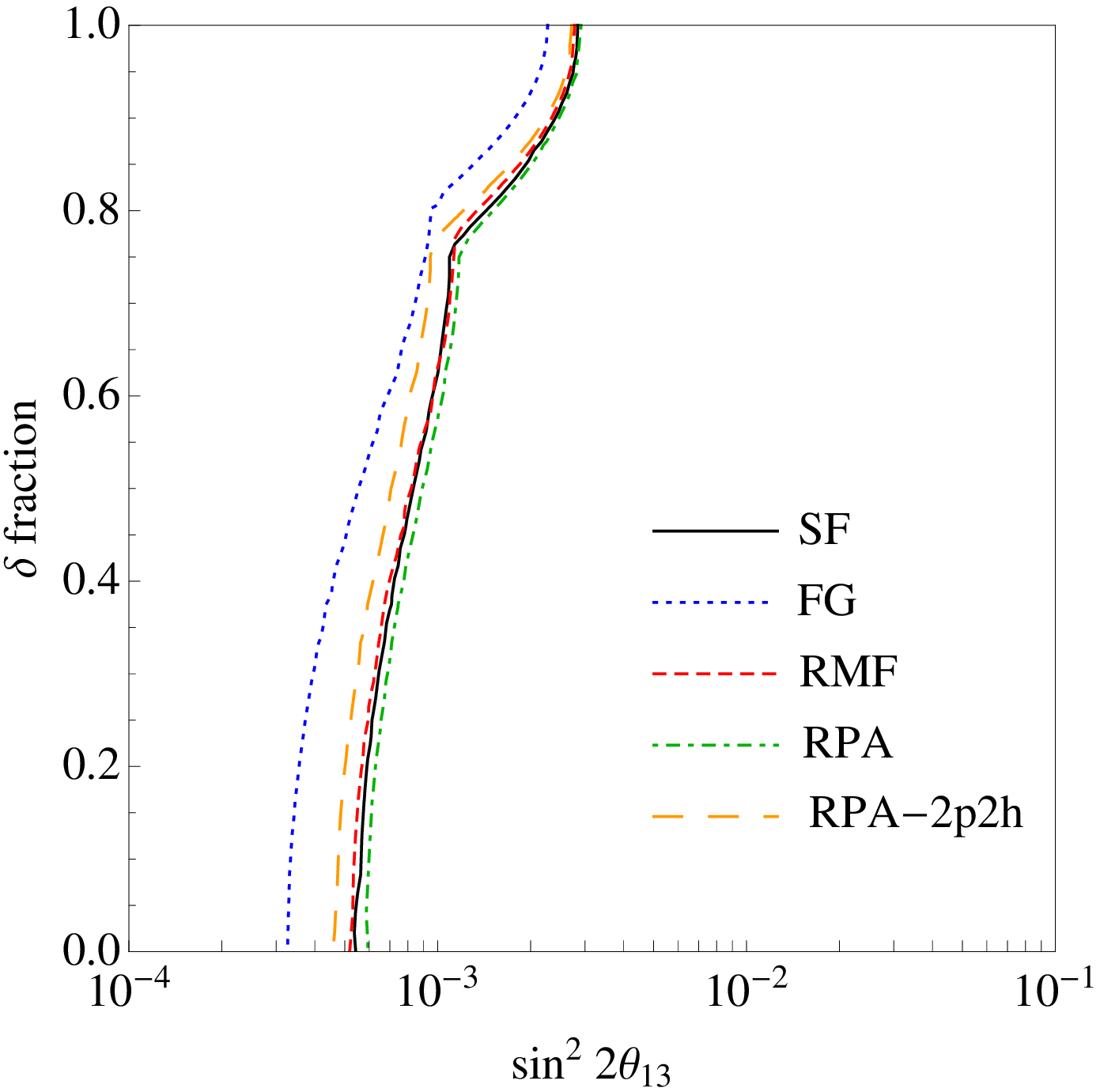}
\caption{\label{fig:disc2} \it  Left panel: $\theta_{13}$ discovery potential in the 
$(\theta_{13},\delta_{CP})$-plane. 
Solid lines refer to the SF model, dotted lines to the RFG, short-dashed lines to the RMF, dot-dashed to RPA and long-dashed to RPA-2p2h.
Right panel: the CP fraction.}
\end{figure}        
The difference in the sensitivity to $\theta_{13}$ can be seen in Fig. \ref{fig:disc2} where        
we show the results in the $(\sin^2 2\theta_{13},\delta_{CP})$-plane. 
In this case the predictions of the  SF, RMF and RPA models differ by up to a factor of $\sim 60\%$ compared to the RFG model for $\delta_{CP}\sim \pm 90^o$ while the difference is less pronounced for $\delta_{CP}\sim 0^o$. The RPA-2p2h results are more similar to the RFG for $\delta_{CP}\sim 0^o$ and to the other models for $\delta_{CP}\sim 180^o$. This is also evident in the right panel where we show the CP fraction.

\begin{figure}[h!]
\centering
\includegraphics[width=0.45\linewidth]{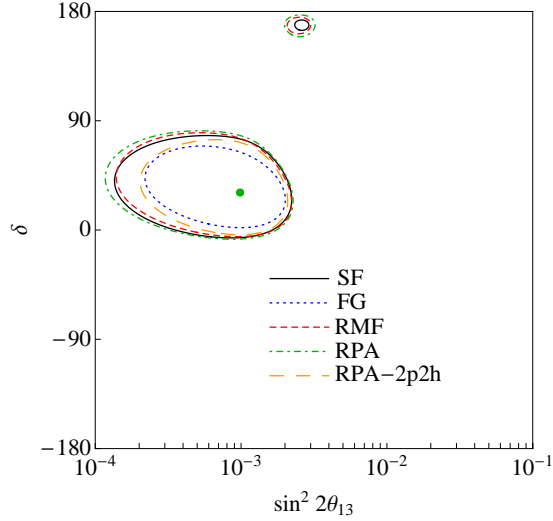}
\caption{\label{fig:patata} \it $90 \%$ CL contour for the input value $(\theta_{13},\delta_{CP})=(0.9^o,30^o)$.
Solid lines refer to the SF model, dotted lines to the RFG, short-dashed lines to the RMF, dot-dashed to RPA and long-dashed to RPA-2p2h.}
\end{figure}

Finally, it is interesting to observe the effect of using different nuclear models also in the simultaneous determination of $\theta_{13}$ and $\delta_{CP}$. We present an example in Fig.\ref{fig:patata} where, for the input value $(\theta_{13},\delta_{CP})=(0.9^o,30^o)$, indicated with a dot, we show the capability of the $\beta$-beam to reconstruct the true values of our observables at $90 \%$ CL.
For the sake of simplicity, we did not include degenerate solutions  coming from our ignorance of the  octant of $\theta_{23}$ and the hierarchy in the neutrino mass ordering but only the so-called ``intrinsic'' one. 
The main feature here is that using the RFG and RPA-2p2h models we are able to reconstruct the 
true values of $\theta_{13}$ and $\delta_{CP}$ within reasonable uncertainties, whereas with the other models
we can only measure two distinct disconnected regions (the fake one around the value of $\theta_{13}$ and $\delta_{CP}\sim 180^o$), which worsen the global sensitivity on those parameters. 
The effects we have mentioned have been generalized to account for the following cases:
\begin{itemize}
\item the value of the axial mass does not fill the gap among the SF approach and the RFG model, at least in the range 
$m_A \in [1,1.2]$ GeV. This is not surprising and has been already the subject of an extensive analysis \cite{Benhar:2009wi};
\item the same sensitivity behaviour as observed in Figs.\ref{fig:disc}-\ref{fig:disc2} has been also seen for a different
nuclear target, namely $^{56}$Fe. This points to the conclusion that the RFG model overestimates the sensitivities to 
$\theta_{13}$ and $\delta_{CP}$ for a vast class of nuclear targets.

\end{itemize}

\section{Conclusion}
\label{concl}
In this letter we have analyzed the impact of different neutrino-nucleus charged current QE cross sections on the forecasted sensitivity of the 
future neutrino facilities to the parameters $\theta_{13}$ and $\delta_{CP}$. We considered five different calculations, 
based on the RFG model (widely used in the MonteCarlo codes) and those based on the Spectral Function approach, on  the Relativistic  
Mean Field  approximation and of the Random Phase Approximation (including multinucleon contributions, also) in the quasi-elastic regime. 
We found that the sensitivities computed from SF, RMF and RPA (without the multinucleon contribution) models are worse than the 
RFG results, for both $\theta_{13}$ and $\delta_{CP}$, by up to a factor 2 when an Oxygen nuclear target is used to compute the event rates. 
To a less extent, this is also true for the RPA-2p2h model, due to the fact that the quasi-elastic cross section is larger than the genuine quasi-elastic cross section. 
These variations with the nuclear models are so large that they cannot be taken into account with the few percent systematic error uncertainty expected from dedicated measurements at near detectors. Indeed, the present uncertainty is much larger and allows for discrepancies between models that can strongly affect the forecasted sensitivity of a given facility, implying that
special care must be adopted when comparing the performance of different facilities.
We have also checked that other nuclear targets produce quantitatively similar results. 
This suggests that the use in MonteCarlo simulations of more refined nuclear models for the neutrino-nucleus interaction is mandatory, especially for 
those facilities whose bulk of events is in the quasi-elastic regime.


\section*{Acknowledgements}
We are strongly indebted with Omar Benhar for providing us the oxygen and iron spectral functions. We also want to thank Maria Barbaro, Juan Antonio Caballero and Jose Udias for providing us the neutrino-nucleus cross sections
for the RMF approach, Marco Martini for the cross sections in the RPA and Magda Ericson and Sergio Palomares Ruiz for useful comments on the manuscript.
D.M. was supported by the Deutsche Forschungsgemeinschaft, contract WI 2639/2-1. E.F.M was supported by the Max-Planck-Institut f\"ur Physik, where this study was commenced, and from CERN, where it was finalized. E.F.M also acknowledges support from the European Community under the European Commission Framework Programme 7 Design Study: EUROnu, Project Number 212372. 

\end{document}